\documentclass[aps,prl,twocolumn,showpacs,groupedaddress]{revtex4}  % for review and submission
\usepackage{graphicx}  % needed for figures
\usepackage{dcolumn}   % needed for some tables
\usepackage{bm}        % for math
\usepackage{amssymb,amsmath}   % for math
\usepackage{amsthm}
\usepackage{subfigure}

\begin{document}

% The following information is for internal review, please remove them for submission
%\leftline{Version xx as of \today}
%\leftline{Primary authors: Joe E. Physics}
%\leftline{To be submitted to (PRL, PRD-RC, PRD, PLB; choose one.)}
%\rightline{Comment to {\tt d0-run2eb-nnn@fnal.gov}}
%\rightline{by xxx, yyy}

% the following line is for submission, including submission to the arXiv!!
%\hspace{5.2in} \mbox{Fermilab-Pub-04/xxx-E}

\title{Emergence of Symmetry in Complex Networks}
% LIST_OF_AUTHORS_R2.TEX                5/16/07
%
\author{Yanghua Xiao$^{1}$}
\author{Momiao Xiong$^{2,3}$}
\author{Wei Wang$^{1}$}
\author{Hui Wang$^{4}$}

\affiliation{\vspace{0.1 in}\vspace{0.1 in}}

\affiliation{$^{1}$Department of Computing and Information
Technology, Fudan University, Shanghai 200433, PR China}

\affiliation{$^{2}$Theoretical Systems Biology Lab , School of Life
Science, Fudan University, Shanghai 200433, PR China}

\affiliation{$^{3}$Human Genetics Center, University of Texas Health
Science Center at Houston, Houston TX 77225, USA}

\affiliation{$^{4}$Business School, University of ShangHai for
Science and Technology, Shanghai 200093, PR China}
  % input Dzero author list
\date{\today}

\begin{abstract}
Many real networks have been found to have a rich degree of
\emph{symmetry}, which is a very important structural property of
complex network, yet has been rarely studied so far. And where does
symmetry comes from has not been explained. To explore the mechanism
underlying symmetry of the networks, we studied statistics of
certain local symmetric motifs, such as symmetric bicliques and
generalized symmetric bicliques, which contribute to local symmetry
of networks. We found that symmetry of complex networks is a
consequence of similar linkage pattern, which means that nodes with
similar degree tend to share similar linkage targets. A improved
version of BA model integrating similar linkage pattern successfully
reproduces the symmetry of real networks, indicating that similar
linkage pattern is the underlying ingredient that responsible for
the emergence of the symmetry in complex networks.

\end{abstract}

 \maketitle

 In the last
decades, we have witnessed the great progress in the complex network
researches
\cite{albertSM,albertWWW,amaral,barabasiBA,kleinberg,uspowergrid,song,newman,motif,Newman20031,Newman2002,Newman20032,Newman2004,ravasz}.
Previous studies have primarily focused on finding the statistical
properties of various networks, such as small world
property\cite{amaral,kleinberg,uspowergrid}; power-law distribution
of vertex degree\cite{barabasiBA}; building block of network
motifs\cite{motif}; assortative mixing\cite{Newman2002};
self-similarity\cite{song}; community structure
\cite{Newman20031,Newman20032,Newman2004}and hierarchical
structure\cite{ravasz} of the network. And based on these
properties, many network models, such as Baraba{\'s}i-Albert
(BA)\cite{barabasiBA} model, Watts-Strogatz model\cite{uspowergrid}
have been proposed to help predict the future evolution of the
network. However, an important property of network structure,
\emph{symmetry}, has been rarely studied.

Concept of symmetry is based on the concept of automorphism of the
graph, which characterizes adjacency invariance to transformation
operation on the node set. Graph has been widely used to represent
systems consisting of components (represented by nodes) as well as
their relation (represented by edges). If two nodes are connected by
an edge they are defined as adjacent nodes. An automorphism acting
on the node set can be viewed as a permutation of the nodes of the
graph preserving the adjacency of the nodes. The set of
automorphisms under the product of permutation forms a group
\cite{Godsil}. In general, a network is considered as
\emph{asymmetric} if its underlying graph contains only an identity
permutation, otherwise, the network is \emph{symmetric}.

It has bee shown that various complex networks have a rich degree of
symmetry  \cite{sym1,sym2}. The fact that large real networks are
symmetric is surprising\cite{sym1}, since 'almost all graphs are
asymmetric'\cite{bollobasRG}\footnote{which could be understood in
the asymptotic way that the proportion of graphs on $n$ vertices
that are asymmetric goes to 1 as n tends to $\infty$}. As an
ubiquitous phenomenon, the existence of symmetry in the real
networks strongly begs an explanation, since existing ingredients,
such as \emph{continuous growth} and \emph{preferential
attachment}\cite{barabasiBA} dominating the construction of the
network structure, are not dedicated to interpret the origination of
symmetry in real networks.

 To explore the origin of symmetry
in real networks, we summarize statistics of the local symmetric
motifs contributing to the symmetry of the real networks, by which
we found that \emph{ similar linkage pattern\footnote{ It may be
confused between \emph{similar linkage pattern} and
\emph{assortative mixing}. Both of the concept focus on the behavior
of those nodes having similar properties. However, assortative
mixing emphasize on inter-linkage between these nodes, while Similar
linkage pattern only require that these nodes share similar linkage
target, whether these nodes are inter-linked is not significant.}},
which means \emph{that nodes having similar property, for example
degree, tend to have similar linkage targets}, is a ubiquitous law
that dominating the construction of structures of a variety of real
networks. For example, in a friendship network, it is widely
believed that persons with similar properties such as educational
background, interest, age, would probably have common friends.

\begin{figure}
\centering \subfigure[] { \label{fig:sym:a}
\includegraphics[scale=1]{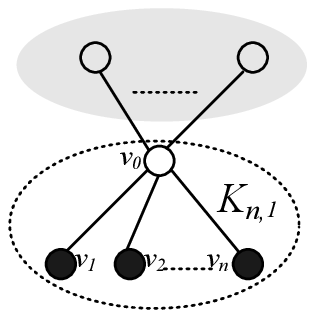}}
\hspace{0.2in}
 \subfigure[]{ \label{fig:sym:b}
\includegraphics[scale=1]{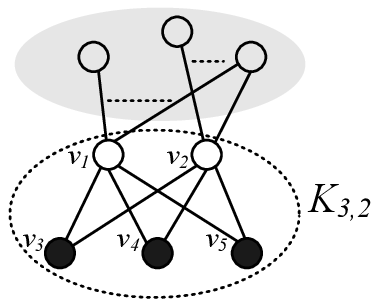}}
\caption{Illustration of symmetric bicliques. Figure (a) shows an
example of $K_{n,1}$, which contributes to the symmetry of the
network with a geometric decomposition factor $n!$; figure (b) shows
an example of $K_{3,2}$, which contributes to the symmetry of the
network with a geometric decomposition factor $S_3$.}
\label{fig:sym} %% label for entire figure
\end{figure}

To show that \emph{similar linkage pattern} is a ubiquitous law that
holds across many structures of real networks, we first summarize
the statistics of \emph{symmetric bicliques} in the real networks,
which is a \emph{induced complete bipartite subgraph}, denoted as
$K_{V_1,V_2}$, in which the degree of vertices in $V_1$ is
conserved\footnote{which means that for each $v\in V_1$,
$d_K(v)=d_G(v)$, where $d_G(v$) is the degree of vertex $v$ in graph
$G$}. Obviously, if graph $G$ contains a symmetric biclique
$K_{V_1,V_2}$, then the automorphism group of $G$, denoted as
$Aut(G)$, will have a corresponding \emph{geometric decomposition
factor}\cite{sym2} $S_n$ with $n=|V_1|$, which indicates that the
size of $Aut(G)$ has a factor of $n!$. Thus, $K_{V_1,V_2}$ becomes a
\emph{local symmetric motifs}\cite{sym2} contributing to the
symmetry of the network. Hence, symmetric bicliques will contribute
to the symmetry of the network. Figure \ref{fig:sym} illustrate two
such bicliques.

 If we do not care about what $V_1$
and $V_2$ are, we also use ${K}_{i,j}$ to denote $K_{V_1,V_2}$,
where $|V_1|=i$ and $|V_j|=j$. And the set consisting of all
${K}_{i,j}$ is denoted as $\mathcal{K}_{i,j}$. Note that
$\mathcal{K}_{1,i}$ does not necessarily contribute to the local
symmetry of the network, hence, in the following discussion, only
$\mathcal{K}_{n,i}$ with $n\geq 2$ has been summarized.

\begin{table*}
\caption{\label{tab:spl_real}Symmetric biclique statistics of a
number of real networks.We measure the size of the networks by the
number of nodes and edges, denoted by N and M, respectively. For
each $i\leq 7$, the statistics of symmetric bicliques contained in
$\mathcal{K}_{n,i}$with $n\geq 2$ is measured. We use a triple tuple
$(S,Min,Max)$ to show the statistics of $\mathcal{K}_{n,i}$, where
$S$ is the number of non-equal substructures in $\mathcal{K}_{n,i}$,
and the $Min, Max$ are the minimal and maximal size of symmetric
bicliques, respectively. If $\mathcal{K}_{n,i}$ is empty, $S=0$, and
$Min$ and $Max$ is not available, denoted as '-'. For some larger
$i$, we also enumerate corresponding statistics of
$\mathcal{K}_{n,i}$.}
\begin{ruledtabular}
\begin{tabular}{ccccccccc}
 &\multicolumn{8}{c}{$\mathcal{K}_{n,i}$with $n\geq 2$} \\
 Network&1&2&3&4&5&6&7& some larger i
\\ \hline
arXiv\cite{arXiv}\protect\footnote{Here, the snapshot at 2006-03 of
HEP--TH (high energy physics theory) citation graph \cite{arXiv} is
used.}
&(135,2,7)&(42,2,4)&(17,2,3)&(13,2,2)&(11,2,2)&(1,2,2)&(2,2,2)&$i=16$,(1,2,2)\\
InternetAS\protect\footnote{Here, the snapshot at
2006-07-10 of CAIDA\cite{caida} is used.}&(916,2,343)&(1057,2,285)&(90,2,25)&(9,2,4)&(2,2,2)&(0,-,-)&(0,-,-)&(0,-,-)\\
BioGrid\cite{biogrid}\\
SAC&(51,2,15)&(7,2,5)&(0,-,-)&(0,-,-)&(0,-,-)&(0,-,-)&(0,-,-)&(0,-,-)\\
MUS&(7,2,44)&(8,2,12)&(4,2,6)&(2,2,2)&(0,-,-)&(1,2,2)&(0,-,-)&(0,-,-)\\
HOM&(366,2,44)&(53,2,12)&(21,2,6)&(5,2,2)&(2,2,2)&(1,2,2)&(0,-,-)&$i=8,10,21$,(1,2,2)\\
DRO&(418,2,40)&(16,2,11)&(6,2,3)&(6,2,3)&(3,2,2)&(0,-,-)&(3,2,3)&$i=8,10,21$,(2,2,2),$i=15,27$,(1,3,3)\\
 &&&&&&&&$i=17,18,19,23,25$ (1,2,2) $i=39$,(1,6,6)\\
CAE&(245,2,47)&(9,2,5)&(1,2,2)&(0,-,-)&(0,-,-)&(0,-,-)&(0,-,-)&(0,-,-)\\
USPowerGrid\cite{uspowergrid}&(137,2,9)&(25,2,3)&(0,-,-)&(1,2,2)&(0,-,-)&(0,-,-)&(0,-,-)&(0,-,-)\\
\end{tabular}
\end{ruledtabular}
\end{table*}

\begin{table}
\caption{\label{tab:spl_rnd}Symmetry biclique statistics for a
variety of ER random networks. For each network tested in Table
\ref{tab:spl_real}, we generate the corresponding ER random networks
with the same size using \emph{PAJEK}\cite{pajek}. Two parameters
are needed, the vertex number $N$ and the average degree $z$ to
ensure the generated random network having the same size with the
corresponding real networks. Similar to the evaluation in Table
\ref{tab:spl_real}, we also measure the statistics for each
$\mathcal{K}_{n,i}$, but only statistics of $\mathcal{K}_{n,1}$ is
available for some random networks. Larger symmetric bicliques
$\mathcal{K}_{n,i}$ with larger $i$, do not exist in these random
networks. }
\begin{ruledtabular}
\begin{tabular}{ccccc}
 Network&N&z&$\mathcal{K}_{n,1}$ with  $n\geq 2$
\\ \hline
arXiv&27770&25.37&(0,-,-)\\
InternetAS&22442&4.06&(62,2,3)\\
BioGrid\\
SAC&5437&26.86&(0,-,-)\\
MUS&218&3.65&(1,2,2)\\
HOM&7522&5.32&(3,2,2)\\
DRO&7528&6.69&(0,-,-)\\
CAE&2780&3.13&(21,2,3)\\
USPowerGrid&4941&1.49&(231,2,3)\\
\end{tabular}
\end{ruledtabular}
\end{table}

\begin{figure}
\includegraphics[scale=0.42]{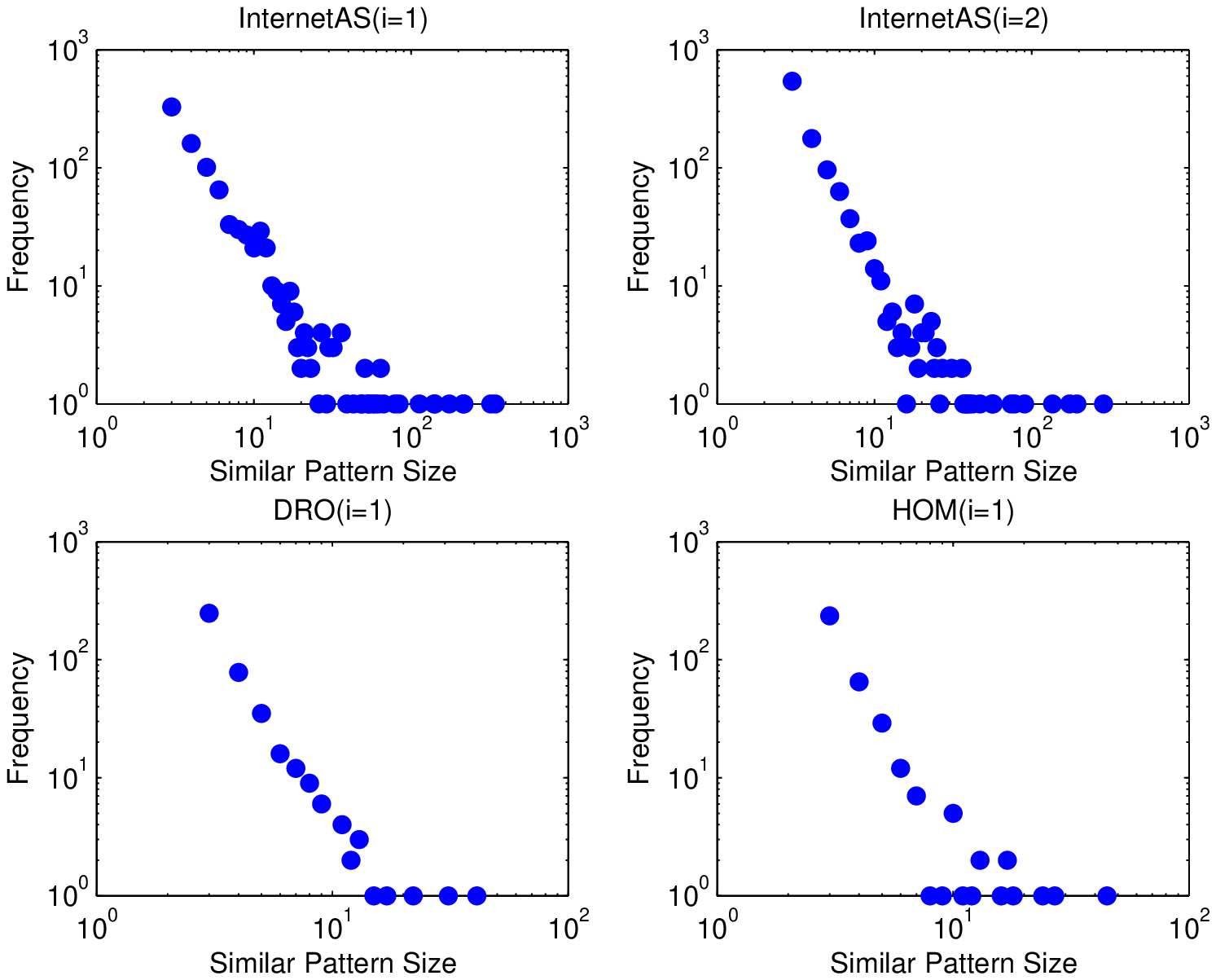}
\caption{\label{fig:dist1}Size distribution of the symmetric
bicliques for real networks. The horizontal axis for each panel is
the size of symmetric bicliques and the vertical axis is the
occurrence frequency of the symmetric bicliques with the
corresponding size. Figure (a) and (b) show the biclique size
distribution of the Internet at autonomous level
\protect\footnote{Here, the snapshot at 2006-07-10 of
CAIDA\cite{caida} is used.} for $\mathcal{K}_{n,1}$ and
$\mathcal{K}_{n,2}$ respectively; Figure (c) shows the biclique size
distribution of $\mathcal{K}_{n,1}$ of \emph{Homo
Sapiens}\cite{biogrid}. Figure (d) shows biclique size distribution
of $\mathcal{K}_{n,1}$ of \emph{Drosophila
melanogaster}\cite{biogrid}.}
\end{figure}

As shown in Table \ref{tab:spl_real} , \emph{similar linkage
pattern} is a universe phenomenon in the process of structure
construction of many real networks including social networks,
biological networks and technological networks. For instance, for
$\mathcal{K}_{n,1}$ on \emph{InternetAS} dataset, there are totally
916 non-disjoint\footnote{Two subgraphs are non-disjoint implies
that the node set of these two graphs are disjoint.} structures,
among which there exists some larger symmetric motifs, e.g. the
maximal motif has 343 nodes in $V_1$. For all the network we tested,
\emph{simple} symmetric motifs such as $\mathcal{K}_{n,1}$ and
$\mathcal{K}_{n,2}$ can be frequently observed. Moreover, for some
networks, such as biogrid network 'DRO', even for some larger $i$,
more complex symmetric motifs of $\mathcal{K}_{n,i}$ do exist.

As shown in Figure \ref{fig:dist1}, among those simple symmetric
motifs with $i=1,2$, the size (n) distributions are right-skewed
with a long tail for larger size, which implies that a number of
larger patterns do exist.

Furthermore, we will show that similar linkage pattern will not
happen in ER \cite{ER} random graphs. As shown in Table
\ref{tab:spl_rnd}, only few randomized networks having the same size
as the corresponding real networks, have symmetric motifs of
$\mathcal{K}_{n,1}$; and no larger motifs $\mathcal{K}_{n,i}$ with
$i\geq2$ exist. Also we found that the number of motifs in
$\mathcal{K}_{n,1}$ of randomized networks is much less than that of
the corresponding real networks with the same size, the complexity
of the motifs are much lower than that of the corresponding real
networks.

The frequent occurrence of complex ${K}_{n,i}$ in real networks and
the unfrequent occurrence of complex ${K}_{n,i}$ in random networks
strongly suggest that there exist some laws dominating the structure
construction process of real networks. Consider the dynamic process
of the network growth. We assume that at some time a new node $v$ is
added to the network, and a symmetric motif $K_{V_1,V_2}$ will
arise. Thus, from the facts we have observed, it's reasonable to
believe that $v$ will attach to the existing nodes under the
principle of \emph{preferentially linking to those nodes to which
other nodes in $V_1$ attach}. Since nodes in $V_1$ have the same
degree, it's reasonable to believe that \emph{nodes having the same
degree will have the same linkage pattern.}

\begin{figure}
\centering
 \subfigure[] { \label{fig:gene-cb:a}
\includegraphics[width=1in]{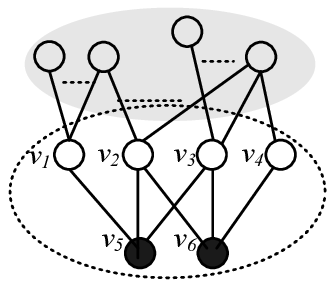}}
\subfigure[] { \label{fig:gene-cb:b}
\includegraphics[width=1in]{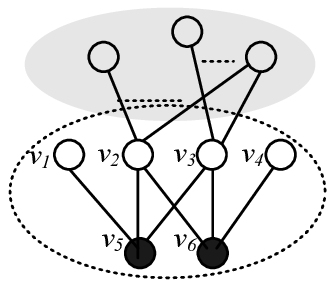}}
\subfigure[] { \label{fig:gene-cb:c}
\includegraphics[width=1in]{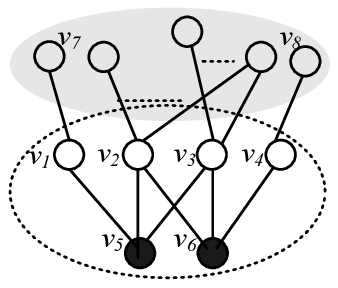}}
\caption{Illustration of non-exact similar linkage pattern. All
nodes in $V_1$, i.e., dark nodes, have the same number of linkages.
However, the linkage target of these nodes are not exactly the same.
Motifs induced by $V_1$ as well as their linkage targets, do not
necessarily contribute to the local symmetry of networks. In Figure
\ref{fig:gene-cb:a}, the subgraph induced by $V=\{v_1,...,v_6\}$
will not lead to any automorphism, while the induced subgraph in
Figure \ref{fig:gene-cb:b} will result in an automorphism
$p=(v_1,v_4)(v_5,v_6)$ and the subgraph induced by
$V=\{v_1,...,v_8\}$ in Figure \ref{fig:gene-cb:c} also contributes
an automorphism $p=(v_1,v_4)(v_5,v_6)(v_7,v_8)$ to the symmetry of
the graph.}
 \label{fig:gene-cb}
\end{figure}

However, as shown in Figure \ref{fig:gene-cb}, in real networks,
nodes having the \emph{same} degree tend to have only \emph{similar}
targets not exactly the same target sets, and these local motifs
exhibiting non-exact similar linkage behavior also have chance to
contribute to the symmetry of the network. Clearly, these local
substructures are the generalization of symmetric bicliques, in the
way that structures constraint of the clique is relaxed from being
complete bipartite to only satisfying that all the vertex of $V_1$
have the same degree in the clique. Thus, this kind of
\emph{generalized symmetric biclique} can be denoted as
$K^d_{V_1,V_2}$, where $d$ is the degree of any vertex in $V_1$; in
some cases that we do not care about $V_2$, $K^d_{V_1}$ or
$K^d_{|V_1|}$ is often used.

In a network, if non-exact similar linkage pattern does make sense,
then nodes with the same degree will tend to share similar linkage
targets. Thus, we need to measure in what extent these nodes share
the same targets. Let $V(m)=\{v:v\in V \text{ and } d(v)=m\}$ be all
nodes with degree $m$, then the linkage targets of these nodes could
be denoted as $V(m)'=\{v':(v,v')\in E \text{ and } v\in V(m)\}$.
Then we could define $\theta_m$ as Equation \ref{equ:eq_theta},
which is the ratio of the actual number of linkage targets of $V(m)$
to the maximal probable number of linkage targets( Maximal set of
linkage targets could be obtained when overlapping of targets of
nodes in $V(m)$ is forbidden.) Obviously, this measure is a key
index that can be used to quantify overlapping ratio of linkage
target of nodes in $V(m)$.
 \begin{equation}
\theta_m=\frac{|V'(m)|}{m|V(m)|} \label{equ:eq_theta}
 \end{equation}

\begin{figure}
\centering
 \includegraphics[scale=0.5]{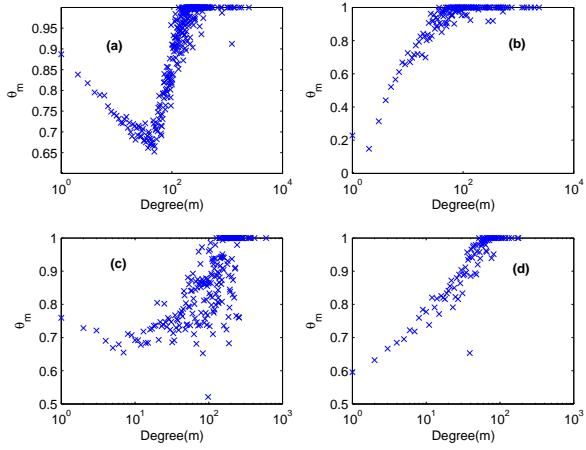}
\caption{Figure(a),(b),(c),(d) show $\theta_m$ distribution of
arXiv, Internet, Biogrid SAC and Biogrid DRO, respectively. }
\label{fig:slp} %% label for entire figure
\end{figure}

Obviously, we have $0<\theta_m\leq1$. If $|V_m|$ is given, we have
$\frac{1}{|V(m)|}\leq\theta_m\leq1$. Note that, the lower $\theta_m$
is, the more frequently similar linkage pattern will happen, while
for $\theta_m$'s closer to $1$, the exact opposite is true. As shown
in Figure \ref{fig:slp}, for small values of degrees, all tested
networks tend to have relative small $\theta_m$, which strongly
suggest that for these real networks, in the process of network
growth, \emph{nodes with the same small degree tend to have the
similar linkage behavior}.

It has been shown in the BA\cite{barabasiBA} model that many real
networks have power law degree distribution, which can be attributed
to two basic ingredients: (1) \emph{Growth} and
(2)\emph{Preferential attachment}. In BA model, new nodes will be
continuously added to the existing networks, and at each time step,
a new node is preferentially attached to fixed number of $m$ (this
number $m$ is referred to as \emph{initial degree} of the newly
added node) existing highly connected nodes. However, ingredients
about symmetry have not been considered in BA model and other
 network generation models. To reproduce \emph{symmetric}
networks with power law degree distribution, we propose a new
network model incorporating similar linkage pattern into BA's two
ingredients. For this purpose, two modifications on BA model's two
principles need to be carried out:
\begin{enumerate}
    \item \emph{Newly added nodes are linked to the existing nodes not only under
 the principle of preferential attachment, but also similar linkage
pattern}. The latter principle implies that newly added node with
initial degree $m$ tends to link to the targets to which existing
nodes with degree $m$ in the network are linked.
    \item \emph{Initial degree $m$ of newly added nodes
 follows a certain distribution instead of being a constant value}.
 In BA and other existing models, initial degree is constant,
while in the following study, we will show that in many real
networks, $m$ follows a certain distribution.
\end{enumerate}

%\textbf{Preferential attachment considering similar linkage pattern.}
The probability, denoted by $\Pi$, that a new node with
initial degree $m$ will be connected to node $v_i$ not only rely on
degree $k_i$ of node $v_i$ but also depends on whether $v_i$ belongs
to $V'_t(m)$. To incorporate the ingredient of similar linkage
pattern into the basic BA model, we need to increase $\Pi(v_i)$ for
those $v_i$ belonging to $V_t'(m)$. Hence, we define parameter
$\alpha$ to control the relative significance of similar linkage
pattern in the formation of network structure. Note that for a given
$m$, $V'_t(m)$ is not necessarily to be non-empty, hence the
probability $\Pi$ would be defined in two cases: when
$V'_t(m)=\emptyset$, then $\Pi$ should be defined as:
\begin{equation}
    \Pi(v_i)=\frac{k_i}{\sum_jk_j}.
\label{eq:pi1}
 \end{equation}
where $k_i$ is the degree of vertex $v_i$; when
$V'_t(m)\neq\emptyset$, $\Pi$ should be defined as :
\begin{equation}
    \Pi(v_i)=
     \begin{cases}
 \alpha\frac{k_i}{\sum_jk_j}+(1-\alpha)\frac{1}{|V'_t(m)|}& \text{if $v_i\in V'_t(m)$},\\
 \alpha\frac{k_i}{\sum_jk_j}& \text{if $v_i\notin V'_t(m)$}.
\end{cases}
\label{eq:pi2}
 \end{equation}, where $\alpha\in(0,1]$.

At some time step $t$, we may have $V'_t(m)=\emptyset$ and $\Pi$
would be reduced to pure preferential attachment in terms of the
value of degree. It will happen frequently in the initial stage of
the network growth in our model. Because the abundance of degree is
limited in the initial stage, whatever the seed network is. For
example, if the seed network contains isolated vertices, only degree
0 could be found, if the seed network is an regular network, such as
a complete network, we could also find only one degree in the
network.

Equation \ref{eq:pi2} has only one parameter $\alpha$ to control the
relative significance of pure preferential attachment and similar
linkage pattern. It's clear that the larger $\alpha$ is, the less
impact of similar linkage pattern on the network will be. When
$\alpha=1$, the model is reduced to the pure preferential attachment
according to vertex degree.

\begin{figure}
\centering
\includegraphics[scale=0.3]{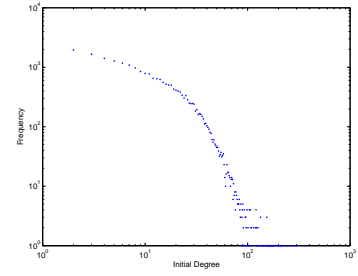}
 \caption{Distribution of initial degree of arXiv dataset  }
\label{fig:initial_degree} %% label for entire figure
\end{figure}

%\textbf{Initial degree following certain distribution.}
Obviously,the basic assumption of BA model is that any node except
for those seed nodes has the same initial degree. However, for some
networks, especially social networks and technique networks, whose
historical data about initial degree of real networks is available,
we can easily find that the initial degree of real networks may be
far away from a fixed value or a value independent of degree. For
example, Figure \ref{fig:initial_degree} shows the distribution of
initial degree of a citation network constructed from \emph{arXiv}
data set. From this figure, we can see that for larger initial
degree, the frequency follows power law distribution rather than a
fixed value.

Assume that we grows the network in the way following the principle
of preferentially attachment with similar linkage pattern. If the
initial degree is constant, then each time a new node was added to
the network, fixed number ($m$) edges would be introduced into the
network. Thus, the local symmetric motifs would concentrate on those
subgraphs with structure closer to $K_i^m$. If $m$ is very larger
than 1, it's contradictory to the above observed fact that the
larger $n$ is, the less frequently $K_i^n$ tends to occur.

Hence, it is necessary to extend the initial degree from a fixed
value to some given distribution. From this perspective, BA model
would be considered as a special case of our model in the way that
$F(m)$ is specified as a constant value.\\

%\textbf{Network Model based on Similar Linkage Pattern.}
The algorithm of the model incorporating the ingredient of
\emph{similar linkage pattern} is the following:

  (1)\emph{Growth}: Starting from a small number($n_0$) of isolated
 nodes, at every time step, we add a new node with $m$ edges that
 link the new node to $m$ different nodes already present in the
 system, where $m$ follows a distribution $F(m)$ and $m\leq\bar{m}$,
 where $\bar{m}$ is the upper bound of the initial degree $m$.

  (2)\emph{Preferential attachment with similar linkage pattern}: The probability $\Pi$ that a
 new node will be connected to node $v_i$ is defined by Equation
 \ref{eq:pi1} and \ref{eq:pi2}.

  The above improved model based on similar linkage pattern needs just three input parameters
 $(n_0,F(m),\alpha)$. For the notational convenience, the model is doted as
 $SLP(n_0,F(m),\alpha)$, where 'SLP' is the abbreviation of 'Similar Linkage
Pattern'.

To test the effect of similar linkage pattern on symmetry of the
networks, we first give some measures of symmetry of networks. The
degree of the symmetry of a graph usually could be quantified by
$\alpha_G=|Aut(G)|$\cite{bollobasMGT}, i.e., the size of the
automorphism group. In order to compare the symmetry of networks
with different sizes, $\beta_G$ has been used to measure the
symmetry in \cite{sym1}, which is defined as:
\begin{equation}
\beta_G=(\alpha_G/N!)^{1/N}
 \label{equ:beta_G}
\end{equation},
where $N$ is the number nodes in the network.  $\beta_G$  measures
the symmetry relative to maximal possible automorphism group of a
graph with $N$ nodes. Another symmetry measure $\gamma_G$ is also
given, which is the ratio of number of nodes in all those
\emph{nontrivial orbits} \footnote{Given automorphism partition
group $(Aut(G),V)$ of graph $G=(V,E)$, we can get a partition
$\mathcal{P}=\{V_1,V_2,...,V_k\}$ in the way that $x$ is equivalent
to $y$ if and only if $\exists g\in Aut(G)$, s.t. $x^g=y$. And each
cell of the partition is called as an orbit of $(Aut(G),V)$. An
orbit is trivial if it only contains a single vertex, otherwise, the
orbit is nontrivial.}(a set of equivalents nodes under automorphism
operation) to the number of all nodes in the network. Specifically,
let $\mathcal{P}=\{V_1,V_2,...,V_k\}$ be the automorphism partition
under the action of $(Aut(G),V)$ on node set $V$, $\gamma_G$ could
be defined as :
\begin{equation}
\gamma_G=\frac{\sum_{\substack{1\leq i\leq k,|V_i|>1}}|V_i|}{N}
\label{equ:gamma_G}
\end{equation}

\begin{figure}
\centering \subfigure[] { \label{fig:slpmodel:a}
\includegraphics[scale=0.22]{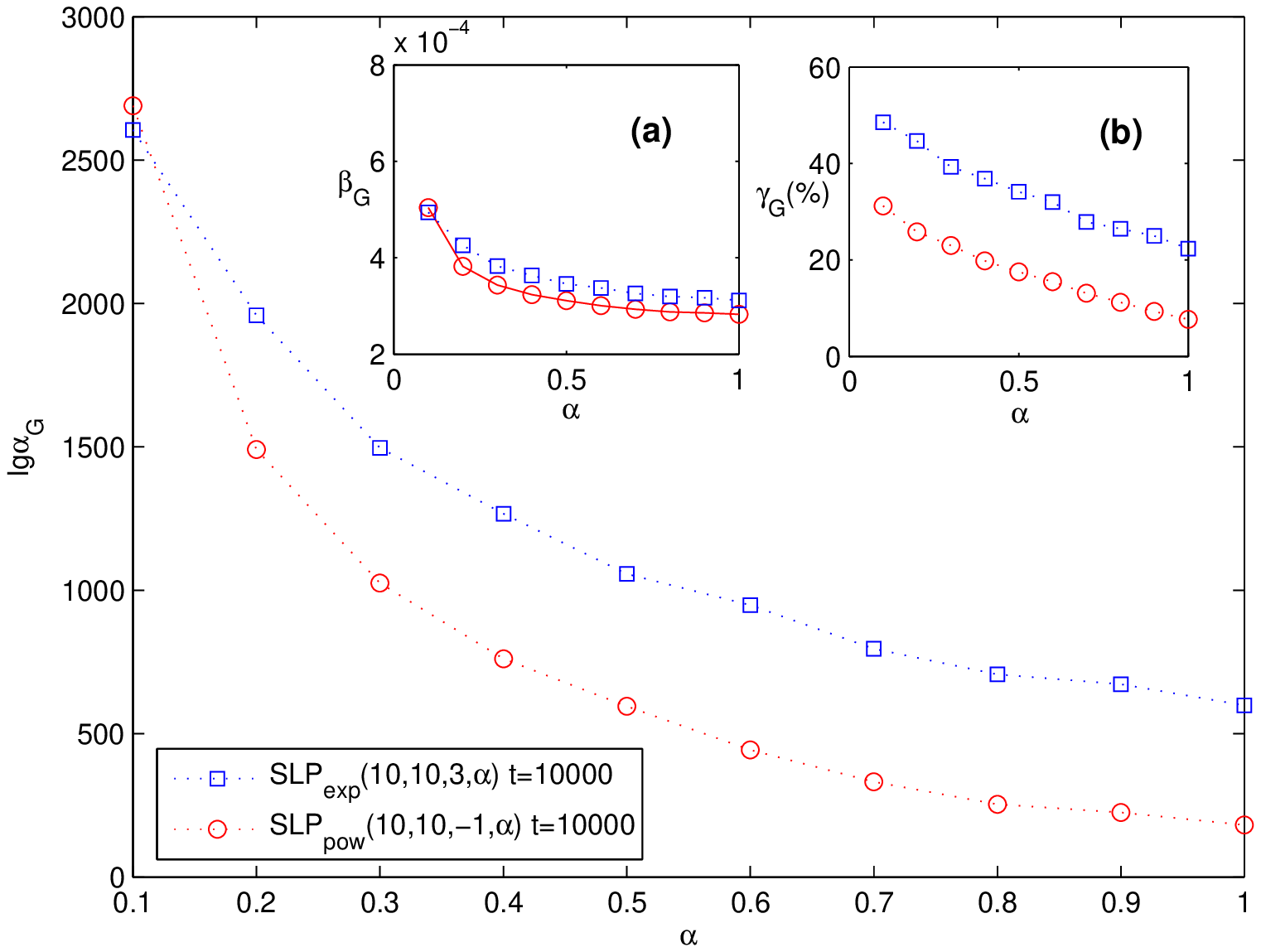}}
\subfigure[] { \label{fig:slpmodel:b}
\includegraphics[scale=0.22]{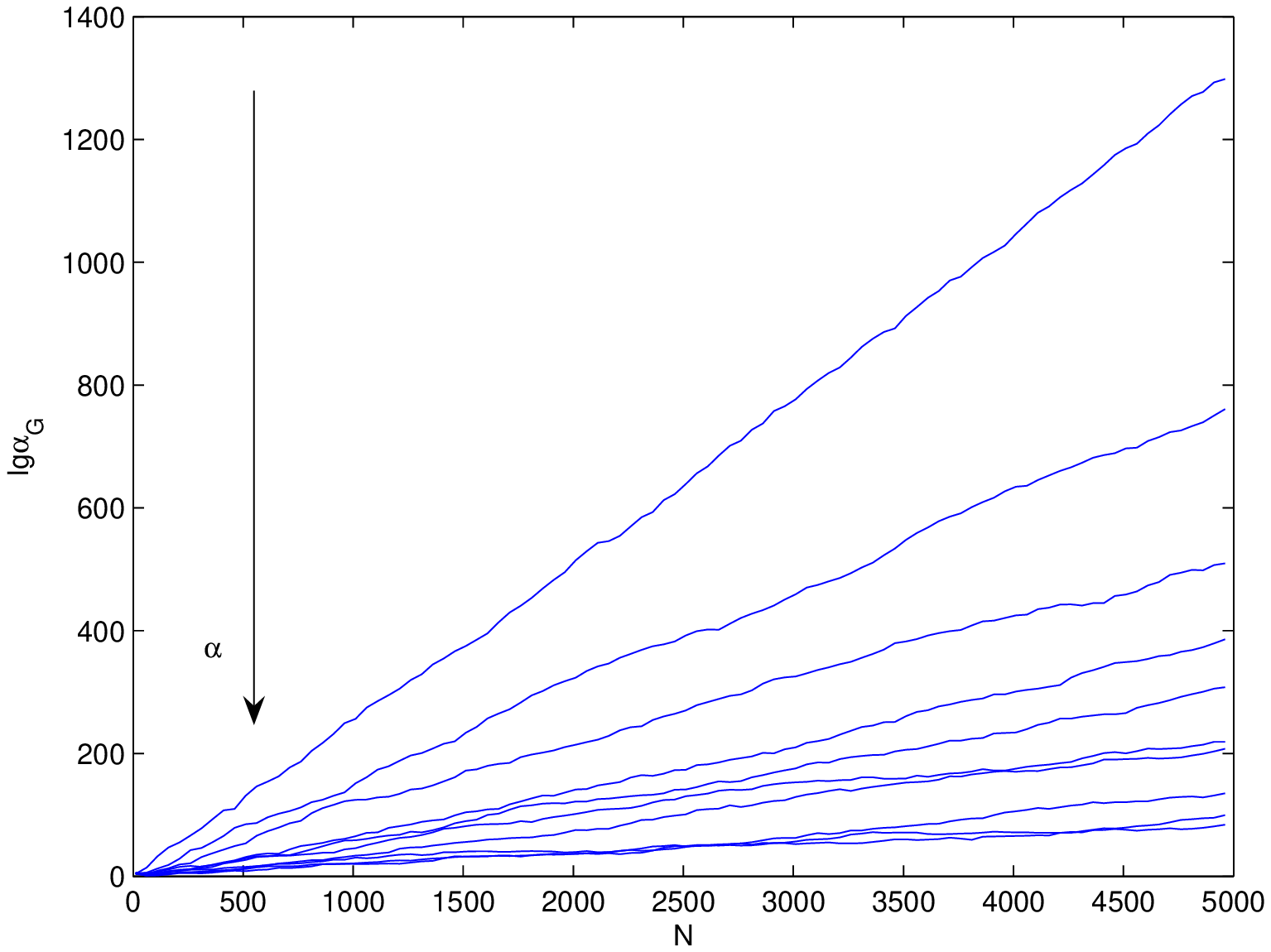}}
\subfigure[] { \label{fig:slpmodel:c}
\includegraphics[scale=0.22]{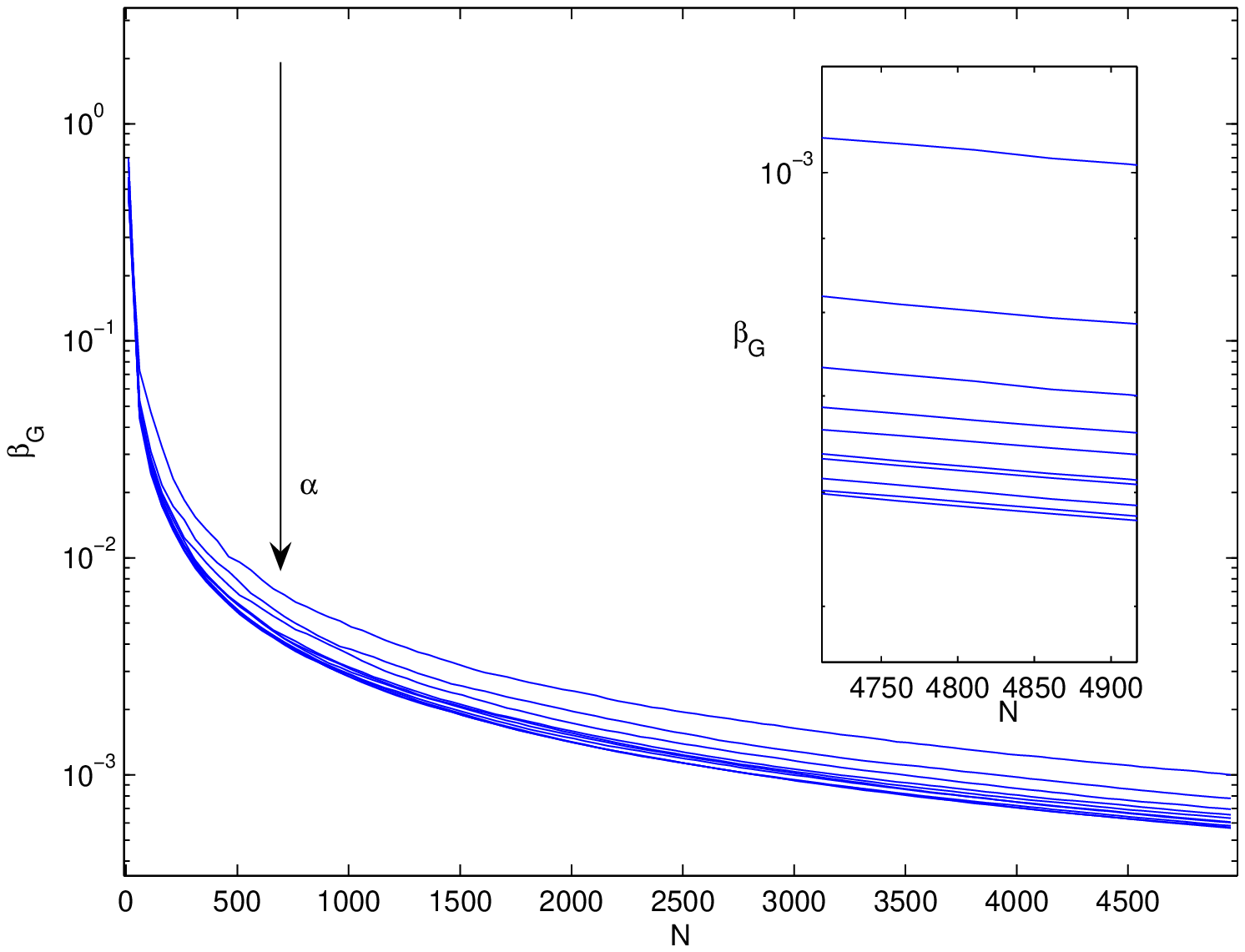}}
\subfigure[] { \label{fig:slpmodel:d}
\includegraphics[scale=0.22]{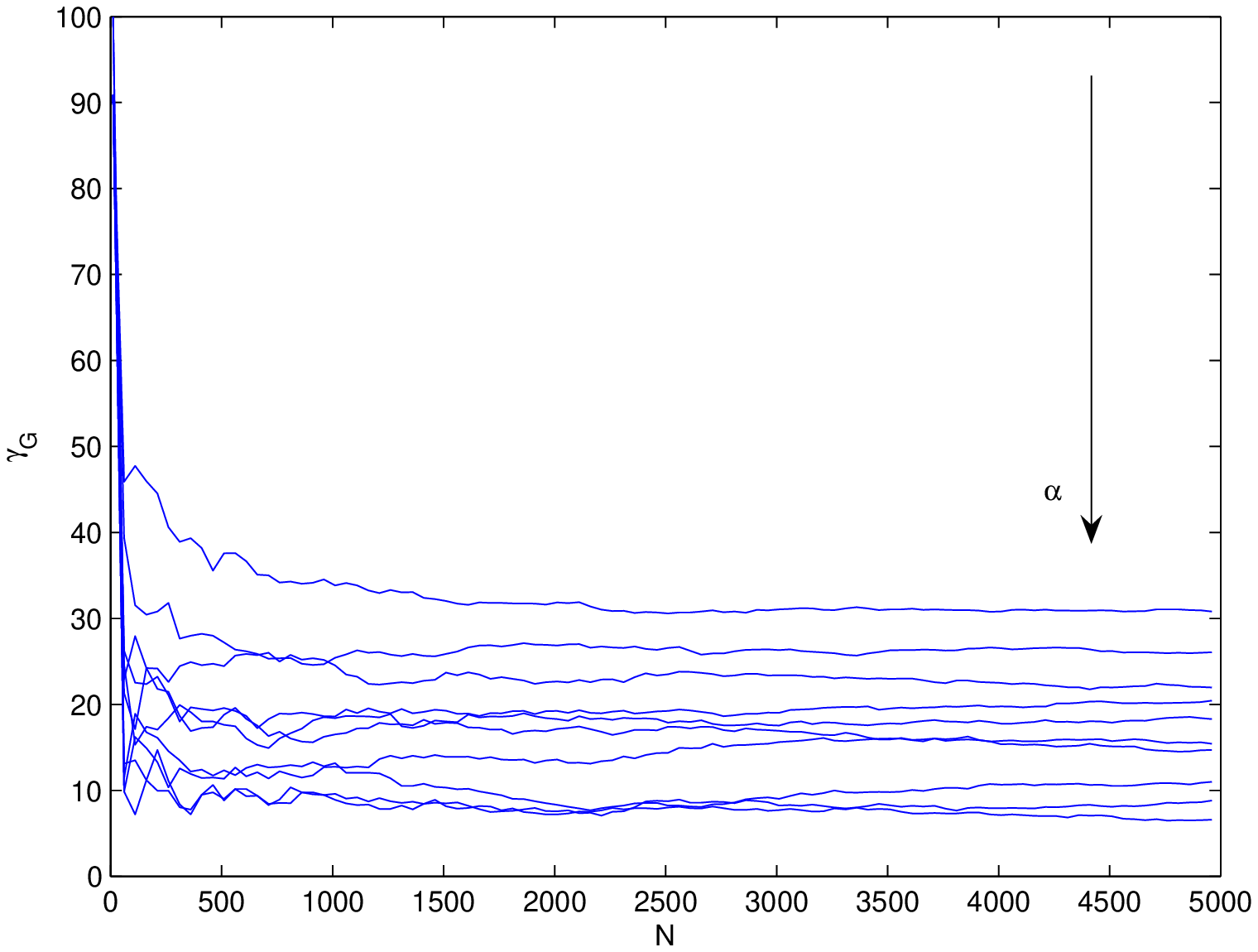}}
\caption{The effect of $\alpha$ and size  on symmetry of networks
generated by the SLP model.(a) Simulation of SLP model with $\alpha$
varying from 0.1 to 1 in increment of 0.1. The horizontal axis of
the figure as well as two insets (a) and (b) is $\alpha$, the
vertical axis of the figure is $\lg\alpha_G$,  the vertical axis of
the inset figure (a) and (b) are $\beta_G$ and $\gamma_G$,
respectively. In the simulations, we use $n_0=10$,$\bar{m}=10$,
$t=10000$ and employ two kinds of initial degree distributions. Blue
square($\square$) shows an exponential distribution
$P(m)=a\gamma^{-m}$ with $\gamma=3$; red circle($\circ$) shows a
power law distribution $P(m)=am^{\gamma}$ with $\gamma=-1$. Figure
(b),(c) and (d) show the growth of symmetry indices including
$\lg\alpha_G$,$\beta_G$ and $\gamma_G(\%)$ of networks generated by
SLP model. In the simulations shown in Figure (b),(c) and (d), a
power law initial degree distribution is employed with $\bar{m}=10$
and $\gamma=-1$, we fix $n_0$ as 10 and vary $\alpha$ from 0.1 to 1(
The arrow shows the direction of increasing $\alpha$). We vary $t$
from 0 to 5000 and capture the snapshots of the network every 50
units of time, thus we could get 100 samples of networks with
linearly increasing sizes. Clearly, for all $\alpha$, the growth of
automorphism group size $\alpha_G$ shows a obvious exponential
trend; and decrease of $\beta_G$ shows an power law trend; and
$\lim_{N\rightarrow\infty}\beta_G=\sigma(\alpha)$ .When $\alpha$
varies from 0.1 to 1, all three symmetry indices decrease. The inset
of Figure (c) shows the amplified local plot, which clearly shows
the fact that $\beta_G$ decrease with the growth of $\alpha$.}
\label{fig:slpmodel} %% label for entire figure
\end{figure}

As shown in Figure \ref{fig:slpmodel:a}, with $\alpha$ varying from
1 to 0.1, i.e., with more influence resulting from similar linkage
pattern exserted on the network construction, the automorphism group
size of networks increase several hundreds of orders of magnitudes.
The inset (a) and (b) of Figure \ref{fig:slpmodel:a} also show that
another two symmetry indices $\beta_G$ and $\gamma_G$ increase with
the decrease of $\alpha$. Such facts also can be observed from the
Figure \ref{fig:slpmodel:b},\ref{fig:slpmodel:c} and
\ref{fig:slpmodel:d}. Hence, it's reasonable to believe that
\emph{similar linkage pattern} is responsible for the emergence of
the symmetry in the networks.

If we remove the ingredient of \emph{similar linkage pattern}, we
will find that ingredient of preferential attachment with initial
degree following a distribution will not necessarily reproduce
symmetry of networks.

\begin{figure}
\centering \subfigure[] { \label{fig:slp_avg:a}
\includegraphics[scale=0.22]{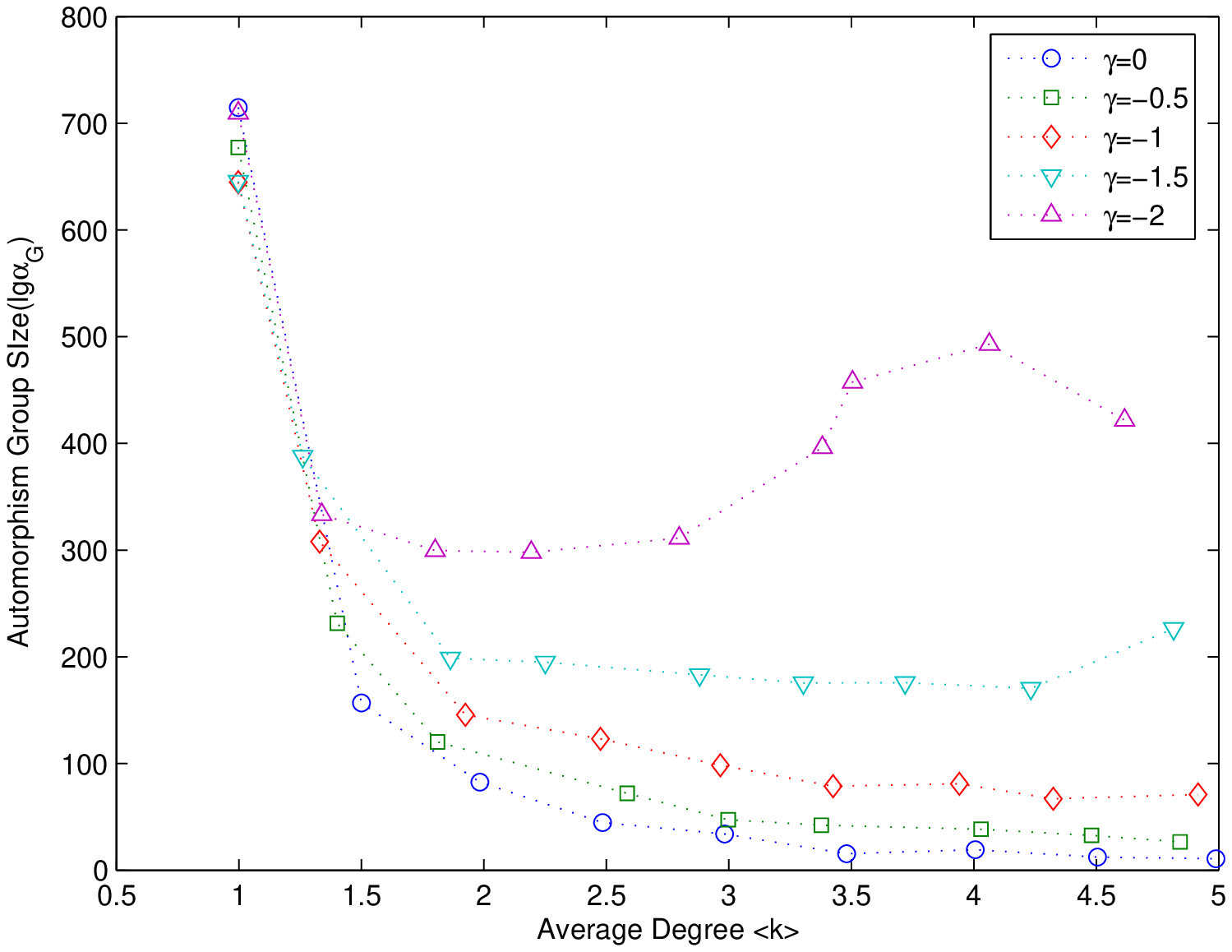}}
\subfigure[] { \label{fig:slp_avg:b}
\includegraphics[scale=0.22]{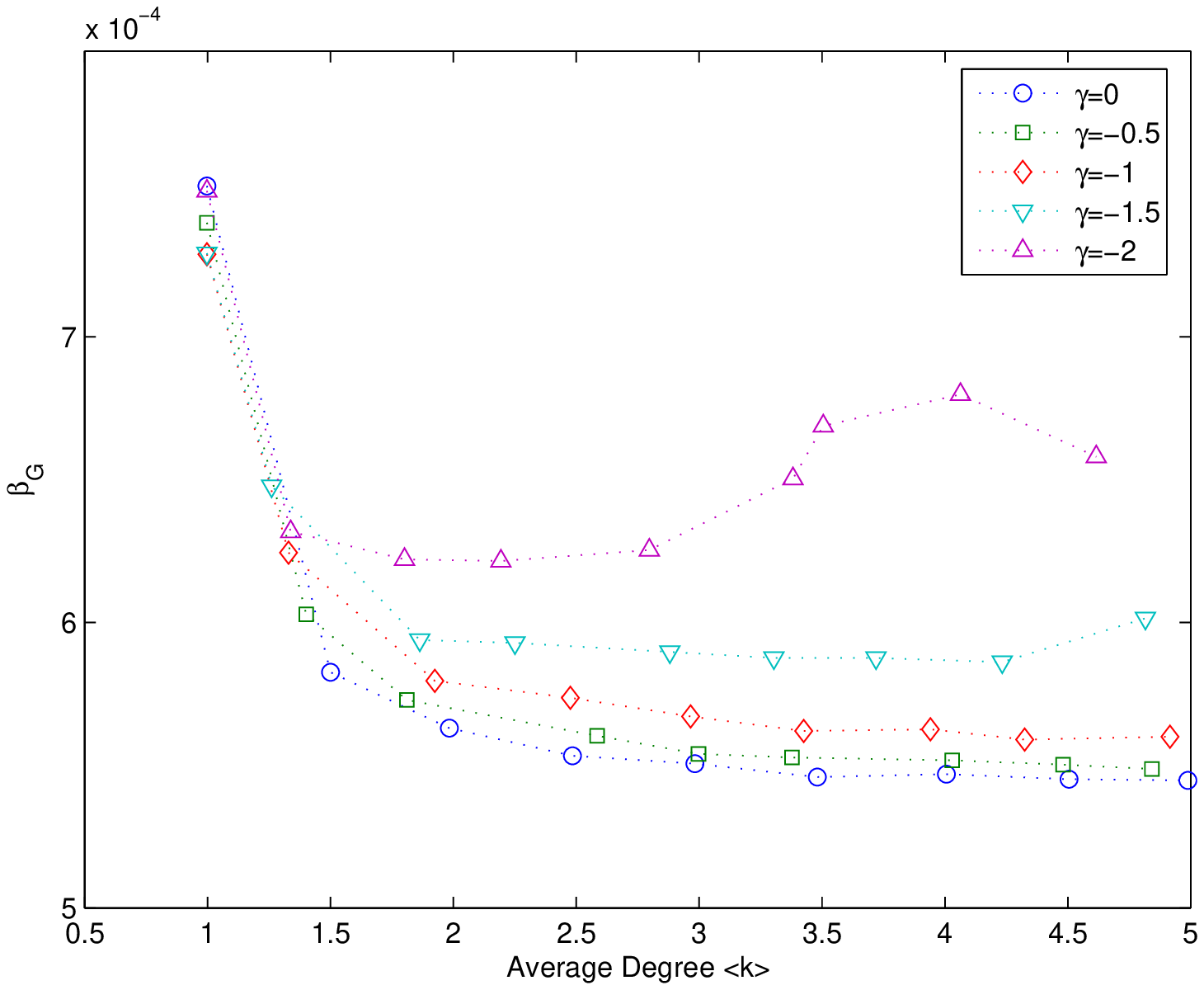}}
\subfigure[] { \label{fig:slp_avg:c}
\includegraphics[scale=0.22]{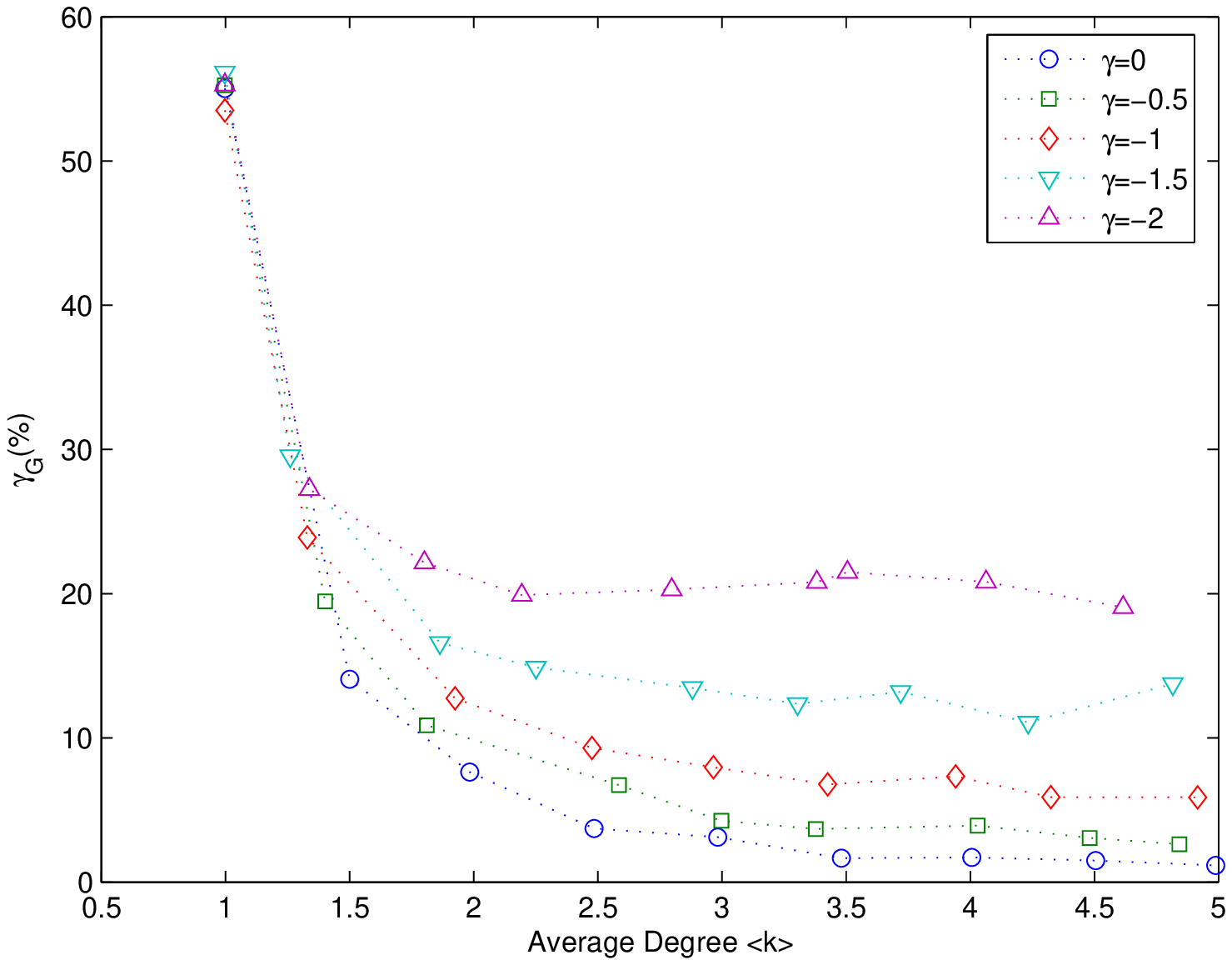}}
\subfigure[] { \label{fig:slp_avg:d}
\includegraphics[scale=0.22]{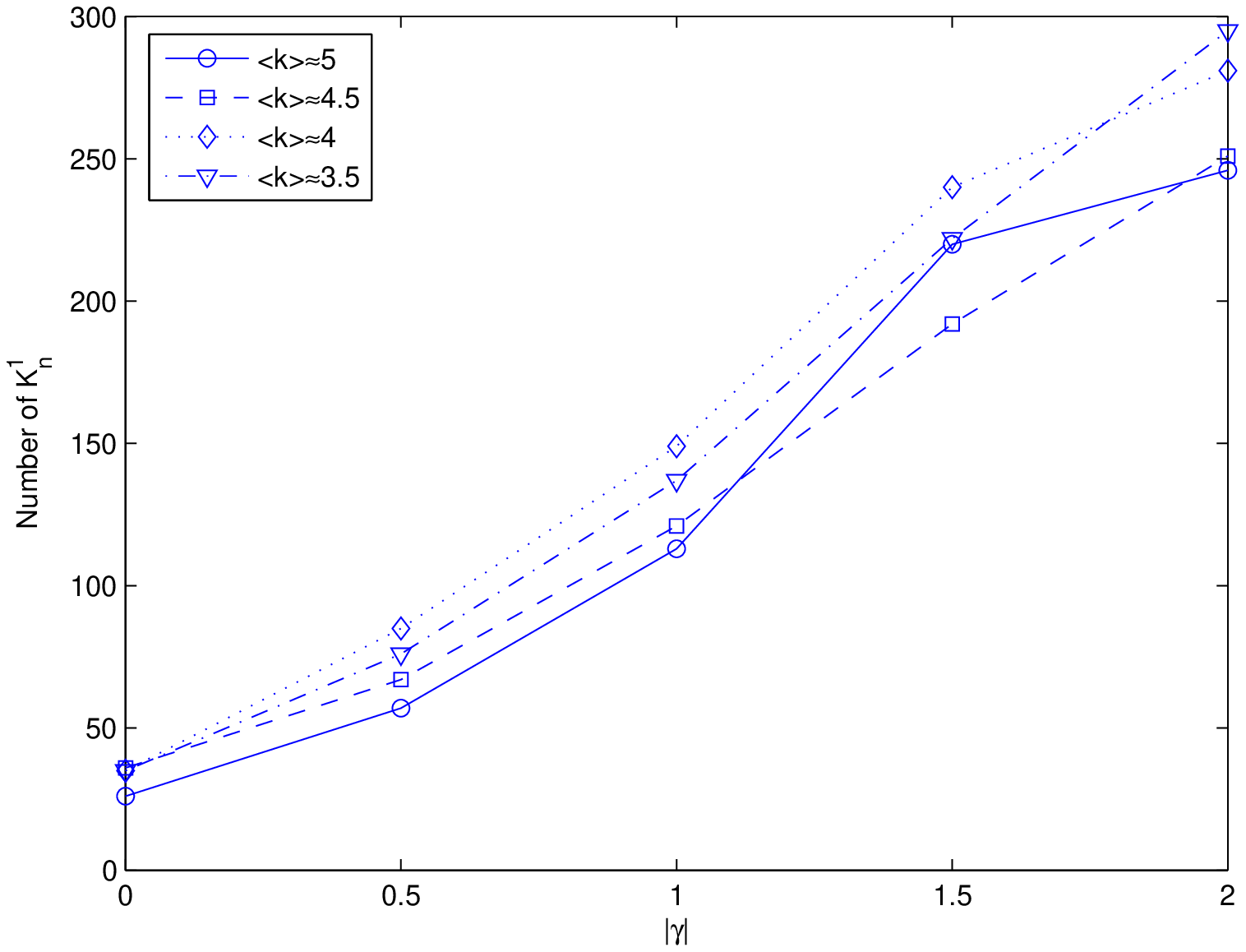}}
\caption{ The effect of average degree $\langle k\rangle$ on
symmetry of networks generated by preferential attachment with
initial degree following a power law distribution
$P(m)=am^{\gamma}$. In this simulation, we use $\alpha=1$ to
eliminate the influence of similar linkage pattern, and other
parameters are set as $n_0=10$, $t=5000$, and
$\gamma=0,-0.5,-1,-1.5,-2$. For each $\gamma$, we vary $\langle
k\rangle$ from 1 to 5 in increments of 0.5.  Figure (a),(b),(c) show
the trend of automorphism group size $\lg\alpha_G$, $\beta_G$ and
$\gamma_G$(\%) with the growth of the average degree of the network,
respectively. It is clear that symmetry of the network will rapidly
(super linearly) decrease to an constant level for less steeper
initial degree distribution. Figure (d) shows the relation between
the number of local structure $K_{n,1}$ and the slop ($|\gamma|$) of
the power law initial degree distribution for average degree as one
of \{5,4.5,4,3.5\}. Parameters in Figure (d) are the same as Figure
(a),(b) and (c). Clearly, with the increase of $|\gamma|$, more
$K_{n,1}$ will occur as the substructures of the network.}
\label{fig:slp_avg} %% label for entire figure
\end{figure}

As shown in Figure \ref{fig:slp_avg}, when average degree $\langle
k\rangle$\footnote{In our study, for the convenience of denotation,
the average degree $\langle k\rangle$ is defined as $\frac{M}{N}$,
where $N$ and $M$ are the numbers of nodes and edges, respectively.
Obviously, $\langle k\rangle$ is a half of the actual average
degree.} is small (close to 1), the network have higher degree of
symmetry. Note that those networks with $\langle k\rangle$ closer to
1 tend to have the structure of \emph{tree} and it is desirable that
tree tends to have higher degree of symmetry. Such result conforms
to the result reported in \cite{sym1} that \emph{BA random} trees
and \emph{uniform random} trees have higher degree of symmetry.

As shown in Figure
\ref{fig:slp_avg:a},\ref{fig:slp_avg:b},\ref{fig:slp_avg:c}, when
 $\langle k\rangle$ increase, the symmetry of the
network rapidly decays to a constant level $\tau(\gamma)$, which is
determined by the slope of the power law distribution. Obviously,
the steeper the initial degree distribution is, the higher the
symmetry level is. When $\gamma=0$, i.e., the slope of the double
log distribution plot is zero, then symmetry of the network rapidly
decays to zero or value close to 0 as $\langle k\rangle$ increases.
However, with the slope becoming steeper, the symmetry of the
network rapidly decays to an approximately constant value
$\tau(\gamma)$ that is far larger than 0 as $\langle k\rangle$
increases. Thus, for steep log-log initial degree distribution,
non-ignored degree of symmetry would be observed.
 Note that steeper initial degree distribution will result in a higher probability of
smaller initial degree $m$, especially $m=1$; as a result, more
\emph{tree-like} symmetry will be found in the structure of the
network. As shown in Figure \ref{fig:slp_avg:d}, the number of
$K_{n,1}$ increases with the growth of $|\gamma|$; as observed in
Table \ref{tab:gamma_tree}, the complexity and the size of $K_{n,1}$
also increase with the slope of the double log initial degree
distribution.

Thus, it's rational to conclude that only preferential attachment
with initial degree following a distribution, will not necessarily
reproduce symmetry of networks. Only in those cases that small
initial degrees have higher probability of occurrences, especially
$m=1$, will produce tree-like symmetry of networks. To reproduce
higher probability of smaller initial degree, we need to decrease
the maximal initial degree or increase the slope of the initial
degree distribution.

\begin{table}
\caption{Statistics of $K_{n,1}$ in some SLP networks with power law
initial degree distribution. All the parameters are set the same as
Figure \ref{fig:slp_avg:d}. In this table, we records the number,
the minimal, the maximal size of the maximal size  of
$\mathcal{K}_{n,1}$ with $\gamma$ as one of \{0,-0.5,-1,-1.5,-2\}
and $\langle k\rangle$ as one of \{3.5,4,4.5,5\}.}
\begin{ruledtabular}
\begin{tabular}{cccccc}
 &&&$\gamma$&&\\
 $\langle k\rangle$&0&-0.5&-1&-1.5&-2
\\ \hline
3.5&(35,2,4)&(76,2,11)&(137,2,10)&(222,2,14)&(295,2,26)\\
4&(35,2,5)&(85,2,5)&(149,2,6)&(240,2,9)&(281,2,41)\\
 4.5&(36,2,4)&(67,2,10)&(121,2,9)&(192,2,18)&(251,2,47)\\
5&(26,2,4)&(57,2,5)&(113,2,6)&(220,2,16)&(246,2,40)\\
\end{tabular}
\end{ruledtabular}
\label{tab:gamma_tree}
\end{table}

In summary, through the statistics of certain local symmetric
motifs, such as (generalized) symmetric bicliques in many real
networks, we found that similar linkage pattern plays an important
role in the origin of symmetry of networks. To incorporate this
ingredient into BA model, we improved BA model in two aspects: (1)
extending the initial degree from a constant value to a
distribution; (2) increasing the linkage probability of those target
nodes. Simulation shows that similar linkage pattern is responsible
for the emergence of symmetry of networks, while preferential
attachment with initial degree following a distribution will only
reproduce tree-like symmetry in some cases.

Extensive existence of similar linkage pattern in real networks
inspires us that behavior of individual nodes is far away from
randomness, which provide us a brand new perspective,
\emph{symmetry}, to understand the self-organization of the complex
systems. From this new viewpoint, in our studies we have found
strong positive correlation between similar linkage pattern and
symmetry of networks, which demonstrates that the emergence of
complexity at macro system level is originated from the simple micro
mechanism of individual component of the network system. Since
symmetry breaking is the basic mechanism underlying the procedure of
the network growth, we believe our studies will be of great value to
help explore the laws dominating symmetry breaking in complex
networks.

\section{\label{sec:level1}Acknowledgement}
The work was supported by the National Natural Science Foundation of
China under Grant No.60303008; the National Grand Fundamental
Research 973 Program of China under Grant No.2005CB321905
   % input acknowledgement

\section{\label{sec:level1}References and notes}

\end{document}